\documentclass[aps,pra,preprint,floatfix,longbibliography, superscriptaddress]{revtex4-1}

\usepackage{pifont}
\usepackage[dvipsnames]{xcolor}
\definecolor{bananayellow}{rgb}{1.0, 0.88, 0.21}
\definecolor{amethyst}{rgb}{0.6, 0.4, 0.8}
\definecolor{ao(english)}{rgb}{0.0, 0.5, 0.0}

\usepackage{float}
\usepackage{epsfig}
\usepackage{bm}
\usepackage[matrix,frame,arrow]{xy}
\usepackage[applemac]{inputenc}
\usepackage[T1]{fontenc}
\usepackage{lmodern}
\usepackage[english]{babel}
\usepackage{amsmath}
\usepackage{ae}
\usepackage{units}
\usepackage{amssymb}
\usepackage{color}
\usepackage{graphicx}
\usepackage{url}
\usepackage{nomencl}
\usepackage{subfigure}
\usepackage{slashed}

\usepackage{fixmath}
\usepackage{booktabs}

\usepackage{xr}
\usepackage[colorlinks]{hyperref}
\hypersetup{%
	plainpages=true,
	breaklinks=true,% not default in dvips mode, so we must specify
	hypertexnames=false,%not ideal, but needed when pagenums duplicate (`i' vs. `1')
	pageanchor=true,
	colorlinks=true,
	linkcolor={blue},
	citecolor={red},
	urlcolor={blue},
	anchorcolor={black}
}

\usepackage{mleftright} %to get rid of the extra space around parentheses with \left and \right: write \mleft, \mright instead 

\newcommand{\ket}[1]{|#1\rangle}

\newcommand{\brakket}[3]{\langle #1 | #2 | #3 \rangle}
\newcommand{\expec}[1]{\langle #1 \rangle}

\newcommand{\comm}[2]{\mleft[ #1, #2 \mright]}
\newcommand{\lind}[1]{\mathcal{D}\mleft[#1\mright]}

\newcommand{\sz}{\hat \sigma_z}
\newcommand{\sx}{\hat \sigma_x}
\newcommand{\sm}{\hat \sigma_-}

\newcommand{\abs}[1]{\mleft|#1\mright|}

\newcommand{\abssq}[1]{\mleft| #1 \mright|^2}

\newcommand{\nn}{\nonumber}

\newcommand{\figref}[1]{\mbox{Fig.~\ref{#1}}}

\newcommand{\secref}[1]{\mbox{Sec.~\ref{#1}}}

\newcommand{\appref}[1]{\mbox{Appendix \ref{#1}}}
\renewcommand{\eqref}[1]{\mbox{Eq.~(\ref{#1})}}

\newcommand{\be}{\begin{equation}}
\newcommand{\ee}{\end{equation}}
\newcommand{\bea}{\begin{eqnarray}}
\newcommand{\eea}{\end{eqnarray}}

\newcommand{\beq}{\begin{eqnarray}}
\newcommand{\eeq}{\end{eqnarray}}

\begin{document}
	
\title{Dissipation and Thermal Noise in Hybrid Quantum Systems\\ in the Ultrastrong Coupling Regime}

\author{Alessio Settineri}
%\email[]{alessio.settineri@unime.it}
\affiliation{Dipartimento di Scienze Matematiche e Informatiche, Scienze Fisiche e Scienze della Terra, Universit\`{a} di Messina, I-98166 Messina, Italy}

\author{Vincenzo Macr\'i}
\affiliation{Theoretical Quantum Physics Laboratory, RIKEN Cluster for Pioneering Research, Wako-shi, Saitama 351-0198, Japan}
		
\author{Alessandro Ridolfo}
\affiliation{Theoretical Quantum Physics Laboratory, RIKEN Cluster for Pioneering Research, Wako-shi, Saitama 351-0198, Japan}
	
\author{Omar Di Stefano}
\affiliation{Theoretical Quantum Physics Laboratory, RIKEN Cluster for Pioneering Research, Wako-shi, Saitama 351-0198, Japan}
		
\author{Anton Frisk Kockum}
\affiliation{Theoretical Quantum Physics Laboratory, RIKEN Cluster for Pioneering Research, Wako-shi, Saitama 351-0198, Japan}
\affiliation{Wallenberg Centre for Quantum Technology, Department of Microtechnology and Nanoscience, Chalmers University of Technology, 412 96 Gothenburg, Sweden}
	
\author{Franco Nori}
\affiliation{Theoretical Quantum Physics Laboratory, RIKEN Cluster for Pioneering Research, Wako-shi, Saitama 351-0198, Japan} 
\affiliation{Physics Department, The University of Michigan, Ann Arbor, Michigan 48109-1040, USA}
		
\author{Salvatore Savasta}
\affiliation{Dipartimento di Scienze Matematiche e Informatiche, Scienze Fisiche e Scienze della Terra, Universit\`{a} di Messina, I-98166 Messina, Italy}
\affiliation{Theoretical Quantum Physics Laboratory, RIKEN Cluster for Pioneering Research, Wako-shi, Saitama 351-0198, Japan}
	
\date{\today}
	
\begin{abstract}
The interaction among the components of a hybrid quantum system is often neglected when considering the coupling of these components to an environment. However, if the interaction strength is large, this approximation leads to unphysical predictions, as has been shown for cavity-QED  and optomechanical systems in the ultrastrong-coupling regime. To deal with these cases, master equations with dissipators retaining the interaction between these components have been derived for the quantum Rabi model and for the standard optomechanical Hamiltonian. In this article, we go beyond these previous derivations and present a general master equation approach for arbitrary hybrid quantum systems interacting with thermal reservoirs. Specifically, our approach can be applied to describe the dynamics of open hybrid systems with harmonic, quasi-harmonic, and anharmonic transitions. We apply our approach to study the influence of temperature on multiphoton vacuum Rabi oscillations in circuit QED. We also analyze the influence of temperature on the conversion of mechanical energy into photon pairs in an optomechanical system, which has been recently described at zero temperature. We compare our results with previous approaches, finding that these sometimes overestimate decoherence rates and understimate excited-state populations.
\end{abstract}
	
\maketitle

\section{Introduction}

According to quantum mechanics, a closed system always displays a reversible evolution. However, no quantum system is completely isolated from its environment; for example, control and readout of a quantum system requires some coupling to the outside world, which leads to dissipation and decoherence (see, e.g.,~\cite{Brune1996,Hofheinz2009,Gu2017, Duffus2017}). Realistic quantum systems should thus be regarded as open, taking into account the coupling to their environments. However, using an exact microscopic approach to include the environment (or reservoir) with its many degrees of freedom is often not feasible. Hence, it is highly desiderable to model open quantum systems using a small number of variables. An adequate description of the time evolution of an open quantum system can be provided by the equation of motion for its density matrix: a quantum master equation~\cite{Breuer2002, Gardiner2004}. Another useful approach is based on the Heisenberg Langevin equation (see, e.g., \cite{Kostin1972,Ford1987,Portolan2008}). Microscopic derivations of master equations start from the Hamiltonian dynamics of the total density matrix (for the system plus the environment). Then, tracing out the reservoir degrees of freedom, and introducing some approximations, a master equation can be derived describing the time evolution of the reduced density matrix only for the system ~\cite{Haake1973}. It turns out that the resulting evolution, in general, is no longer unitary, and the open quantum system evolves into mixed states (see, e.g., \cite{shammah2018}).

A hybrid quantum system combines two or more physical components or subsystems~\cite{Wallquist2009, Xiang2013, Kurizki2015}, with the goal of exploiting the advantages and strengths of the different systems in order to explore new phenomena and potentially bring about new quantum technologies. An important requirement for the realization of a functional hybrid quantum system is the ability to transfer, with high fidelity, quantum states and properties between its different components. Specifically, the effective coupling rate between the subsystems must be large enough to allow quantum state transfers between them within the shortest coherence time of the two subsystems \cite{Kurizki2015}. This interaction regime is usually called the strong-coupling regime \cite{Haroche2013}. Cavity quantum electrodynamics (QED) in the strong-coupling regime has demonstrated great capability and potential for the control and manipulation of quantum states \cite{Haroche2013,Xiang2013,Gu2017}. Further increasing the coupling strength, a hybrid quantum system enters the ultrastrong-coupling (USC) regime when the interaction rate becomes comparable to the transition frequency of at least one of the subsystems~\cite{Gu2017,Kockum2018,Forn-Diaz2018}.

It has been shown that USC can give rise to several new interesting physical effects~\cite{DeLiberato2007, Ashhab2010, Casanova2010, Beaudoin2011, Ridolfo2012, Stassi2013, Garziano2013, DeLiberato2014, Sanchez-Burillo2014, Garziano2015, Garziano2016, Cirio2016, Kockum2017a, Kockum2017, Stassi2017, Qin2018,Ridolfo2018}. Ultrastrong coupling has been achieved in a variety of cavity-QED and other hybrid condensed-matter systems, including semiconductor polaritons in quantum wells~\cite{Anappara2009, Gunter2009, Todorov2010, Geiser2012, Askenazi2014}, superconducting quantum circuits~\cite{Forn-Diaz2010,Niemczyk2010, Baust2016,Forn-Diaz2016,Bosman2017,Yoshihara2017a,Yoshihara2017,Yoshihara2017a, Langford2017, Braumuller2017, Magazzu2017, Yoshihara2017b,PuertasMartinez2018,   Forn-Diaz2017, Yoshihara2017, Chen2017, Bosman2017,  PuertasMartinez2018}, a terahertz metamaterial coupled to the cyclotron resonance of a two-dimensional electron gas (2DEG)~\cite{Muravev2011, Scalari2012, Maissen2014, Zhang2016a, Bayer2017}, organic molecules~\cite{Schwartz2011, Kena-Cohen2013, Gambino2014, Gubbin2014, Mazzeo2014, George2016a}, and in an optomechanical system where a plasmonic picocavity was coupled to vibrations in a molecule~\cite{Benz2016}. In particular, in the case of superconducting quantum circuits, it is possible to reach the USC regime with even just a single artificial atom coupling to an electromagnetic resonator~\cite{Devoret2007,Bourassa2009,Niemczyk2010,Forn-Diaz2010,Forn-Diaz2017,Chen2017}. Recently, coupling rates exceeding the transition frequencies of the components (deep-strong-coupling regime~\cite{Casanova2010}) have been obtained in both a circuit-QED setup~\cite{Yoshihara2017, Yoshihara2017b} and with a 2DEG~\cite{Bayer2017}.

Although the Hamiltonian of a coupled light-matter system contains the so-called counter-rotating terms, allowing the simultaneous creation or annihilation of an excitation in both the matter system and the cavity mode, these terms can be safely neglected for small coupling rates, {if the components interact resonantly or almost resonantly}. However, when the coupling strength becomes a significant fraction of the cavity frequency (or of the emitter's transition frequency), this often-invoked rotating-wave approximation (RWA) is no longer applicable and the anti-resonant terms in the interaction Hamiltonian significantly change the standard cavity-QED physics \cite{Anappara2009}. For example, the number of excitations in the cavity-emitter system is no longer conserved~\cite{Kockum2017a}, even in the absence of drives and dissipation, and the system states become dressed by the presence of virtual excitations \cite{DiStefano2017}. {It has also been demonstrated \cite{Sank2016} that counter-rotating terms can induce anomalous qubit transitions (which do not conserve the excitation number) in a superconducting qubit-resonator system detuned from resonance.}

\begin{figure}
\centering
\includegraphics[width = \linewidth]{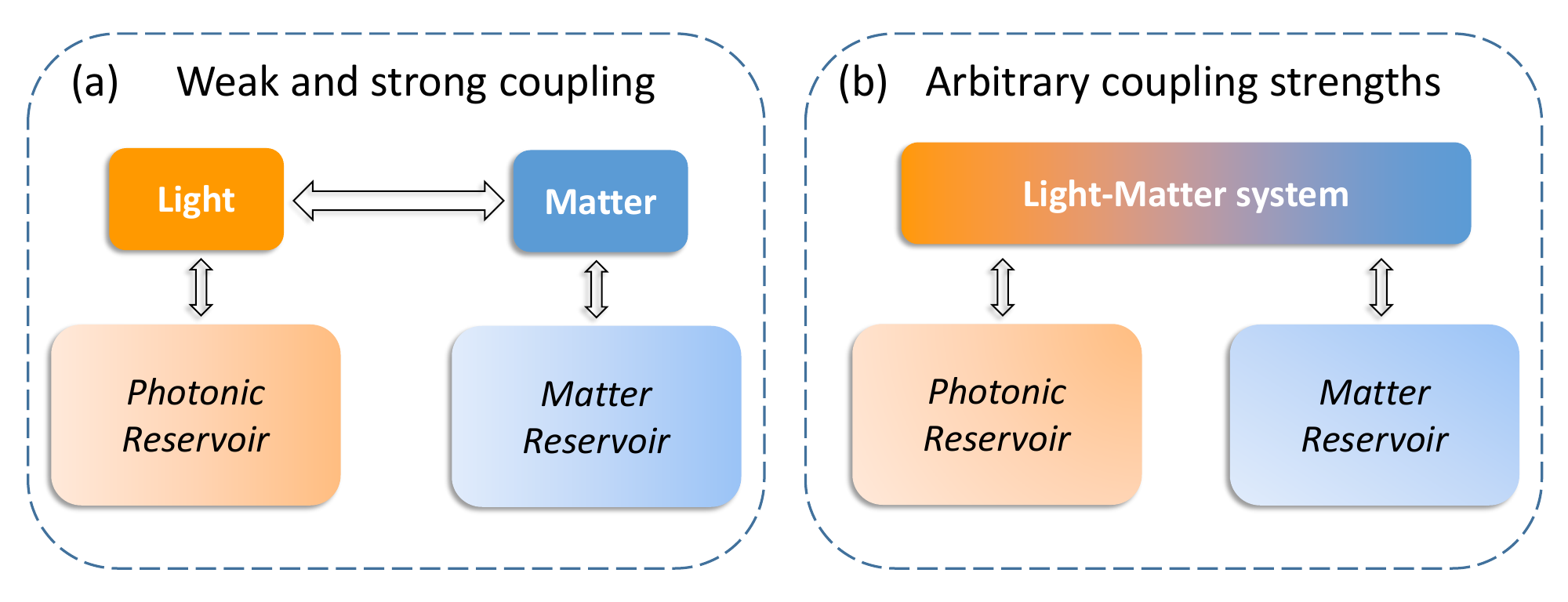}
\caption{(a) The master-equation approach valid in the weak- and strong-coupling regimes. The light-matter coupling is neglected while deriving the dissipators. (b) The master-equation approach considering the light-matter coupling. As the coupling strength between the two subsystems increases, it becomes necessary to treat dissipation effects including the coupling between the subsystems. This can be done by developing the system operators describing the coupling to the reservoirs in the eigenbasis of the coupled light-matter system. 
\label{fig:MasterEqCouplingDiagrams}}
\end{figure}

When deriving the master equation for a hybrid quantum system, the interaction between the subsystems is usually neglected when considering their coupling to the environment [see \figref{fig:MasterEqCouplingDiagrams}(a)]. This results in the standard quantum-optical master equation~\cite{Breuer2002, Gardiner2004} (see \secref{sec:StandardMasterEq}). This procedure works well in the weak-coupling regime, and can also be safely applied in the strong-coupling regime, when the density of states of the reservoirs and the system-bath interaction strengths are approximately flat (frequency independent) on the scale of the energy-level splittings induced by the interaction between the subsystems. However, it has been shown that when the light-matter interaction increases up to the breakdown of the RWA, this approach leads to unphysical predictions, e.g., excitations in the system even at zero temperature~\cite{Beaudoin2011}. A closely related problem arising in the USC regime is the failure of standard input-output theory~\cite{Ciuti2006,Ridolfo2012, Ridolfo2013, Garziano2013, Garziano2017}, which predicts an unphysical output of photons when the hybrid quantum system is in its ground state.

%-----------------------------------------------%
%-----------------------------------------------%

In order to overcome the problems in the description of dissipation of cavity-QED systems in the USC regime, a master equation taking into account the non-Markovian nature of the baths has been developed~\cite{DeLiberato2009}. Furthermore, Ref.~\cite{Beaudoin2011} showed that a master equation working properly in the USC regime of cavity QED can be obtained by including the light-matter coupling in the derivation of dissipative terms of the master equation [see \figref{fig:MasterEqCouplingDiagrams}(b)]. This approach does not require the introduction of  non-Markovian baths. The decoherence rates entering the modified master equation instead depend on the bath noise spectrum evaluated at the dressed transition frequencies of the light-matter system. Since this modified master equation is obtained after a post-trace RWA, it can only be applied to nonlinear interacting quantum systems with anharmonicity larger than the transition linewidths. This prevents the application of this approach to cavity-QED systems in the \textcolor{red}{USC} dispersive regime (see \secref{sec:CircuitQEDBeyondRWA}), and to other hybrid quantum systems displaying a coexistence of harmonic (or quasi-harmonic) and anharmonic transitions, e.g., optomechanical systems. 
In order to describe the losses through the mirror of a cavity embedding matters, a master equation of a non-Lindblad form was also derived \cite{Bamba2016}.
For optomechanical systems in the USC regime, an analogous \textit{dressed-state} master-equation approach has been developed~\cite{Hu2015}, but it also has limitations (see \secref{sec:OptomechBeyondRWA}). A zero-temperature master equation able to describe systems with both anharmonic and (quasi-) harmonic transitions has been introduced to study a cavity-QED system in the USC and dispersive regimes~\cite{Ma2015}. However, a finite-temperature master equation is an essential tool for a precise analysis of experimental results, which, to some degree,  are always affected by thermal noise. {A master equation without the post-trace RWA has been derived to describe a general spin-boson problem mapped into a finite-temperature Rabi model in ultra-strong coupling in Ref.~\cite{Iles2014}.} 

The main purpose of this article is to provide a general approach for the description of dissipation in arbitrary hybrid quantum systems with arbitrary coupling strengths between its components. We do this by presenting a generalized master equation able to describe systems with both harmonic and anharmonic transitions, also valid for non-zero-temperature reservoirs. The only key assumption in our derivation is a weak system-bath interaction, such that the usual second-order Born approximation can be applied (recently, different approaches where this assumption can be relaxed, have been developed in Refs.~\cite{Iles2016,Strasberg2016,DeLiberato2017, Zueco2018}.

In particular, we decompose the system operators in terms of the dressed states of the hybrid quantum system and derive the master equation without performing the usual secular approximation. Finally, we take care of possible numerical instabilities due to the presence of fast oscillating terms.

The outline of this article is as follows: We begin in \secref{sec:ME} by briefly reviewing the standard quantum-optical master equation (\secref{sec:StandardMasterEq}) and the dressed master equation for anharmonic systems (\secref{sec:DressedME}). Section~\ref{sec:GeneralizedME} is devoted to the presentation of a non-Lindblad generalized master equation, able to overcome the limitations of the dressed approach of \secref{sec:DressedME} and to take into account non-zero-temperature reservoirs. We also give a suitable solution for some numerical stability problems of our generalized master equation. In Secs.~\ref{sec:CircuitQEDBeyondRWA} and \ref{sec:OptomechBeyondRWA}, we apply this generalized master equation to calculate the dynamics of a circuit-QED system and an optomechanical system, respectively, at non-zero temperatures, comparing the obtained results with the standard approaches used previously. We conclude in \secref{sec:Conclusions}. In \appref{app:DerivationDressedME}, we present more details for the derivation of the generalized dressed master equation.

%%%%%%%%%%%%%%%%%%%%%%%%%%%%%%%%%%%%%%%%%%%%%%%%%

\section{Master equations}
\label{sec:ME}

In this section, we introduce dissipation for hybrid quantum systems following three different approaches. We start with the standard master equation, generally used for the description of open systems in quantum optics. Then we introduce the dressed master equation \cite{Beaudoin2011}. Finally, we consider a generalized dressed approach, able to describe the dissipation of  hybrid quantum systems with arbitrary coupling strength, valid for systems displaying harmonic, quasi-harmonic, and anharmonic transitions, while also considering non zero-temperature reservoirs.

We begin by considering a generic system consisting of $N$ interacting components or subsystems. Each $i$th component is weakly coupled to an independent bath, modelled as a collection of quantum harmonic oscillators, described by the free Hamiltonian ($\hbar=1$ throughout this article)
\be
\hat H_{\rm B}^{(i)} = \sum_l \nu_l \hat b_{i, l}^\dag \hat b_{i, l} \, ,
\ee
where $\hat b_{i, l}$ ($\hat b_{i, l}^\dag$) are bosonic annihilation (creation) operators for the $l${\rm{th}} bath mode with frequency $\nu_l$ of the $i${\rm{th}} reservoir. The system-bath (denoted by the subscript {\rm SB}) interaction Hamiltonian is given by
\be
\hat H_{\rm SB} = \sum_{i, l} \alpha_{i, l} \mleft( \hat s_i + \hat s_i^\dag \mright) \mleft( \hat b_{i, l} + \hat b_{i, l}^\dag \mright) \, ,
\label{Hsb}
\ee
where $\hat s_i$ ($\hat s_i^\dag$) are annihilation (creation) operators of the $i${\rm{th}} subsystem, mediating the interaction with the reservoirs. We denoted the coupling strength of the $i${\rm{th}} subsystem to the bath mode $l$ of the $i${\rm{th}} reservoir by $\alpha_{i, l}$. In the interaction picture, the system-bath interaction Hamiltonian takes the form
\be
\hat{\tilde H}_{\rm SB} = \sum_{i, l} \alpha_{i, l} e^{\imath \hat H_S t} \mleft( \hat s_i + \hat s_i^\dag \mright) e^{-\imath \hat H_S t} \mleft( \hat b_{i, l} e^{- \imath \nu_{i, l} t} + \hat b_{i, l}^\dag e^{\imath \nu_{i, l} t} \mright) \, ,
\label{hisb}
\ee
where $\hat H_S$ is the system Hamiltonian and $\imath$ is the imaginary unit.

%%%%%%%%%%%%%%%%%%%%%%%%%

\subsection{Standard Master Equation}
\label{sec:StandardMasterEq}

In the standard approach, the components or subsystems are assumed to be independent while obtaining the dissipation. The coupling between the components is afterwards introduced in the system Hamiltonian. This leads to the Schr\"odinger-picture standard master equation
\be
\dot{\hat \rho} = - \imath \comm{\hat H_{\rm S}}{\hat \rho} + \mathcal{L}_{\rm bare} \hat \rho \, ,
\label{MEQstd}
\ee
where $\hat \rho$ is the density matrix of the system and
\be
\mathcal{L}_{\rm bare} \hat \rho = \sum_i \mleft\{ \gamma_i \mleft[ 1 + n \mleft( \omega_i , T_i \mright) \mright] \lind{\hat s_i} \hat \rho + \gamma_i n \mleft( \omega_i, T_i \mright) \lind{\hat s^\dag_i} \hat \rho \mright\} \, ,
\label{Lstd}
\ee
with the generic dissipator
\be
\lind{\hat O} \hat \rho = \frac{1}{2} \mleft( 2 \hat O \hat \rho \hat O^\dag - \hat \rho \hat O^\dag \hat O - \hat O^\dag \hat O \hat \rho \mright) \, .
\ee
In \eqref{Lstd}, the $\gamma_i$'s describe the leakage rates and $n(\omega_i, T_i)$ is the average thermal population of the $i${\rm th} reservoir at temperature $T_i$ and the frequency $\omega_i$ at which $\hat s_i$ rotates in the interaction picture. Pure dephasing effects can be included by adding to \eqref{Lstd} the additional term $(\gamma_{\phi_i} / 2) \lind{\hat d_i} \hat \rho$, where $\hat d_i$ are system operators that do not change the energy of the system, and $\gamma_{\phi_i}$ are the pure dephasing rates. 

The master equation provided in \eqref{MEQstd} can be used to describe many cavity- and circuit-QED experiments in the weak- and strong-coupling regimes~\cite{Gu2017,Gardiner2004,Breuer2002}. However, it has been shown that when the coupling between the components or subsystems increases beyond the point where the RWA is applicable, this approach leads to {\it{unphysical}} predictions, e.g., production of excitations in the system even at zero temperature~\cite{Beaudoin2011}.

%%%%%%%%%%%%%%%%%%%%%%%%%

\subsection{Master equations in the dressed picture}
\label{sec:DressedME}
\subsubsection{Master equation for anharmonic systems}
In order to overcome the limitations of the standard approach, Ref.~\cite{Beaudoin2011} developed a dressed master equation, taking into account the coupling between all the components of the system. They also considered that transitions in the hybrid system occur between dressed eigenstates, not between the eigenstates of the free Hamiltonians of the components.
In the following, we briefly show some key points of the dressed master equation derivation. We first express the system Hamiltonian in the dressed basis of its energy eigenstates. We then switch to the interaction picture, writing the system operators as
\bea
\hat{\tilde S}_i (t) &=& \sum_{j,k>j} C_{jk} |j\rangle \langle k| e^{\imath\Delta_{jk}t}\, ,
\label{dressedop}
\eea
with
\bea
C_{jk}&=&\langle j|(\hat{s}_i + \hat{s}^\dag_i )|k\rangle  \, , \\
\Delta_{jk}&=&E_{j}-E_{k} \, ,
\eea
and the reservoir operators as
\be
\hat{\tilde B} (t) = \sum_{i, l} \alpha_{i, l} \hat b_{i, l} e^{-\imath \nu_l t} \, ,
\label{bt}
\ee
In this way, the system operators $\hat s_i$ are expressed as a sum over transition operators $|j\rangle \langle k|$, which cause transitions (with frequency $\Delta_{jk}$) between eigenstates of the hybrid quantum system $\{|j\rangle,|k\rangle \}$.
Note that `` $\tilde{}$ '' identifies the operators in the interaction picture.
With these new dressed operators, \eqref{hisb} can be split into two parts; one each for the dressed system operators with positive and negative frequencies:
\be
\hat{\tilde H}_{\rm SB} = \sum_i \mleft\{ \hat{\tilde S}_i (t) \hat{\tilde B}_i^\dag (t) + \hat{\tilde S}_i^\dag (t) \hat{\tilde B}_i (t)\mright\} \, .
\label{HsbsplitBlais}
\ee
Note that, as shown in Ref.~\cite{Beaudoin2011}, the fast oscillating terms $\hat S_i^\dag(t)\hat B_i^\dag(t)$ and $\hat S_i(t) \hat B_i(t)$ have been dropped by an initial RWA and the diagonal terms arising from degenerate transitions with $j=k$ are neglected considering a system displaying parity symmetry (in this case $C_{jj}=0$). 
By following the standard procedure~\cite{Breuer2002} (second-order Born approximation, Markov approximation, assuming reservoirs with a continuum of frequencies and performing the secular approximation), as shown in detail in Ref.~\cite{Beaudoin2011}, for this simplified version of \eqref{hisb}, we obtain a dressed master equation that in the Schr\"odinger pincture can be written as,
\be
\dot{\hat \rho} = - \imath \comm{\hat H_S}{\hat \rho} + \mathcal{L}_{\rm dressed} \hat \rho \, ,
\ee
with the Lindbladian superoperator
\be
\mathcal{L}_{\rm dressed} \hat \rho =\sum_{i}\sum_{j, k < j} \mleft\{  \Gamma^{jk}_i n (\Delta_{jk}, T_i) \lind{|j\rangle \langle k|} \hat \rho +  \Gamma^{jk}_i \mleft[ 1 + n (\Delta_{jk}, T_i) \mright] \lind{|k\rangle \langle j|} \hat \rho \mright\} \, ,
\label{lind1}
\ee
where the thermal populations are ($k_B = 1$ throughout this article)
\be
n (\Delta_{jk}, T_i) = \mleft[\exp{\{\Delta_{jk} / T_i\}} - 1 \mright]^{-1}
\ee
and the damping rates are
\be
\Gamma^{jk}_i = 2 \pi g_i (\Delta_{jk}) \abssq{\alpha_i (\Delta_{jk})}\abssq{C_{jk}} \, .
\ee
with $g (\Delta_{jk})$ being the reservoir density of states and $\alpha (\Delta_{jk})$ the system-reservoir coupling strength.

As shown by several studies~\cite{Beaudoin2011, Ridolfo2012, Ridolfo2013, Stassi2013, Altintas2013, Garziano2013, Pagel2015, Cirio2016, LeBoite2016, Stassi2018}, the Lindbladian in \eqref{lind1} can correctly describe the dynamics of anharmonic cavity-QED systems in the USC regime. At $T=0$, rather than exciting the system, the dissipators give relaxation to the true dressed ground state. At $T \neq 0$, these dissipators correctly describe the relaxation to the thermal-equilibrium density matrix for the interacting system~\cite{Ridolfo2013}. However, because of the secular approximation used in the derivation of \eqref{lind1}, this standard approach is {\it not} able to describe dissipation or decoherence in open quantum systems with mixed harmonic-anharmonic or quasi-harmonic spectra~\cite{Beaudoin2011}, e.g, for cavity QED in the dispersive regime and cavity optomechanics.

%%%%%%%%%%%%%

\subsection{Generalized master equation}
\label{sec:GeneralizedME}

\subsubsection{Derivation}

In this section, we extend the previous treatment in order to derive a generalized dressed master equation able to describe both harmonic and mixed harmonic-anharmonic systems coupled to non-zero-temperature reservoirs. Moreover, the present derivation is not limited to systems with parity symmetry.

We start expressing the system Hamiltonian in the dressed basis of its energy eigenstates. We then switch to the interaction picture, writing the system operators as 
\bea
\hat{\tilde S}_i (t) &=& \sum_{\epsilon' - \epsilon = \omega} \hat \Pi (\epsilon) \mleft( \hat s_i + \hat s^\dag_i \mright) \hat \Pi (\epsilon') e^{-\imath \omega t} = \sum_{\epsilon' - \epsilon = \omega} \hat S_i (\omega) e^{-\imath \omega t}\, ,
\eea
and the reservoir operators as in \eqref{bt},
%
%\bea
%\hat{\tilde B} (t) = \sum_{i, l} \alpha_{i, l} \hat b_{i, l} e^{-i \nu_l t} \, ,
%\label{B}
%\eea
%
labelling the eigenvalues of $\hat H_S$ by $\epsilon$ and denoting the projectors onto the respective eigenspaces by $\hat \Pi (\epsilon)\equiv |\epsilon\rangle \langle \epsilon|$. Recall that the symbol \\`` $\tilde{}$ '' identifies interaction-picture operators. In this way, the system operators $\hat s_i$ are expressed as a sum over transition operators, which cause transitions (with transition frequency $\omega$) between energy eigenstates of the hybrid quantum system. For $\omega > 0$, $\hat S_i (\omega)$ is a positive-frequency operator that takes the system from an eigenstate with higher energy to one with lower energy. Conversely, for $\omega < 0$, $\hat S_i (\omega)$ is a negative-frequency operator which produces a transition to a higher-energy eigenstate. In the following, to emphasize these properties, we introduce the notation
\bea
\hat S^{(+)}_i (\omega) &=& \hat S_i (\omega) \hspace{1 cm} {\rm for} \,\, \omega > 0 \, , \label{eq:SPlus}\nn \\
\hat S^{(-)}_i(\omega) &=& \hat S_i (-\omega) \hspace{0.68 cm} {\rm for} \,\, \omega > 0 \, ,\\
\hat S^{(0)}_i &=& \hat S_i (\omega)  \hspace{1 cm} {\rm for} \,\, \omega = 0 \, ,\nn \label{eq:S0}
\eea
With these new dressed operators, \eqref{hisb} can be re-written in a way that makes it easy to derive the Born-Markov master equation for the system:
\be
\hat{\tilde H}_{\rm SB} = \sum_i \hat{\tilde S}_i (t) \mleft[ \hat{\tilde B}_i^\dag (t) +\hat{\tilde B}_i(t) \mright] \, .
\label{HSB1}
\ee

Following the standard procedure (see \appref{app:DerivationDressedME}) the generalized dressed master equation can be obtained evaluating the double integrals in \eqref{rhop1} of \appref{app:DerivationDressedME} without assuming parity symmetry, and evaluating the two integrals without introducing the secular approximation $\omega = \omega'$. In this case, we obtain a Liouvillian superoperator $\cal L$ that, considering all the different subsystems, in the Schr\"odinger picture, can be written in the general form
\bea
\mathcal L_{\rm gme} \hat \rho &=& \frac{1}{2} \sum_i \sum_{\omega, \omega'} \bigg\{ \Gamma_i (-\omega') n (-\omega', T_i) \mleft[ \hat S_i (\omega') \hat \rho (t) \hat S_i (\omega) - \hat S_i (\omega) \hat S_i (\omega') \hat \rho (t) \mright] \nn \\
&&+ \Gamma_i (\omega) n (\omega, T_i) \mleft[ \hat S_i (\omega') \hat \rho (t) \hat S_i (\omega) - \hat \rho (t) \hat S_i (\omega) \hat S_i (\omega') \mright] \nn \\
&&+ \Gamma_i (\omega) [n (\omega, T_i) + 1] \mleft[ \hat S_i (\omega) \hat \rho (t) \hat S_i (\omega') - \hat S_i (\omega') \hat S_i (\omega) \hat \rho (t) \mright] \nn \\
&&+ \Gamma_i (-\omega') [n (-\omega', T_i) + 1] \mleft[ \hat S_i (\omega) \hat \rho (t) \hat S_i (\omega') - \hat \rho (t) \hat S_i (\omega') \hat S_i (\omega) \mright] \bigg\}
%\mathcal{L}_{\rm gme} \hat \rho &=& \sum_{i} \frac{1}{2} \Big\{ \Gamma_i (\omega') n (\omega', T_i) \mleft[ \hat S_i^{(-)} (\omega') \hat \rho \hat S^{(+)}_i (\omega) - \hat S^{(+)}_i (\omega) \hat S_i^{(-)} (\omega') \hat \rho \mright] \nn \\
%&&+ \Gamma_i (\omega) \mleft[ n (\omega, T_i) + 1 \mright] \mleft[ \hat S^{(+)}_i (\omega) \hat \rho \hat S_i^{(-)} (\omega') - \hat S_i^{(-)} (\omega') \hat S^{(+)}_i (\omega) \hat \rho \mright] \nn \\
%&&+ \Gamma_i (\omega) n(\omega, T_i) \mleft[ \hat S_i^{(-)} (\omega') \hat \rho \hat S^{(+)}_i (\omega) - \hat \rho \hat S^{(+)}_i (\omega) \hat S_i^{(-)} (\omega') \mright] \nn \\
%&&+ \Gamma_i (\omega') \mleft[ n (\omega', T_i) + 1 \mright] \mleft[ \hat S^{(+)}_i (\omega) \hat \rho \hat S_i^{(-)} (\omega') - \hat \rho \hat S_i^{(-)} (\omega') \hat S^{(+)}_i (\omega) \mright] \nn \\
%&&+ \sum_i \Omega (T_i) \lind{\hat S^{(0)}_i} \Big\} \, ,
\label{liou}
\eea
where
\be
\Gamma_i (\omega) = 2 \pi g_i (\omega) \abssq{\alpha_i (\omega)} \, ,
\ee
and ``gme'' refers to generalized master equation.
%
%Law in Ref.~[] provided a new master equation obtained without the secular approximation but considering only the case of zero temperature reservoir

Equation~(\ref{liou}) contains several terms since both the transition frequencies $\omega$ and $\omega'$ can be positive, negative and zero, although both $\Gamma_i(\omega)$ and $n(\omega,T_i)$ are non-zero for positive frequencies only. Moreover, only a few of these terms are relevant in order to correctly describe the system dynamics. Indeed, the terms with oscillation frequencies significantly larger than the damping rates  $\Gamma_{i}$ of the system provide negligible contributions when integrating the master equation. Equation~(\ref{liou}) also contains terms with $\omega' = \omega = 0$, originating from diagonal transition operators or, more generally, operators describing zero-frequency transitions. These terms give rise to additional pure dephasing contributions. Note that these terms can be regarded as a generalization of those appearing in the master equation for optomechanical systems in the USC regime~\cite{Hu2015}.

Expanding \eqref{liou}, we obtain terms oscillating at frequencies $\pm (\omega'\pm \omega)$ arising from products of $\hat S^{(-)}_i$ and $\hat S^{(+)}_i$. We also obtain terms oscillating at frequencies $- \omega'$, $+\omega$ arising from products of $\hat S^{(-)}_i$ or $\hat S^{(+)}_i$ with $\hat S^{(0)}_i$, and non-oscillating terms arising from products between zero-frequency operators $\hat S^{(0)}_i$. Moreover, considering a system with well separated energy levels ($\omega \gg \Gamma_i$), the terms oscillating at $\pm (\omega+\omega')$, $+\omega$ and $-\omega'$ can be considered as rapidly oscillating and can be neglected. Including only those terms providing non-negligible contributions to the dynamics, the Liouvillian in \eqref{liou} can be written as
\bea
\mathcal{L}_{\rm gme} \hat \rho &=& \frac{1}{2}\sum_i\sum_{(\omega,\omega')>0} \bigg\{ \Gamma_i (\omega') n (\omega', T_i) \mleft[ \hat S^{(-)}_i (\omega') \hat \rho (t) \hat S^{(+)}_i (\omega) - \hat S^{(+)}_i (\omega) \hat S^{(-)}_i (\omega') \hat \rho (t) \mright] \nn \\
&&+ \Gamma_i (\omega) n (\omega, T_i) \mleft[ \hat S^{(-)}_i (\omega') \hat \rho (t) \hat S^{(+)}_i (\omega) - \hat \rho (t)\hat S^{(+)}_i (\omega) \hat S^{(-)}_i (\omega') \mright] \nn \\
&&+ \Gamma_i (\omega) [n (\omega, T_i) + 1] \mleft[ \hat S^{(+)}_i (\omega) \hat \rho (t) \hat S^{(-)}_i (\omega') - \hat S^{(-)}_i (\omega') \hat S^{(+)}_i (\omega) \hat \rho (t) \mright] \nn \\
&&+ \Gamma_i (\omega') [n (\omega', T_i) + 1] \mleft[ \hat S^{(+)}_i (\omega) \hat \rho (t) \hat S^{(-)}_i (\omega') - \hat \rho (t) \hat S^{(-)}_i (\omega') \hat S^{(+)}_i (\omega) \mright] \nn \\
&&+ {\rm \Omega}^{+}_i (T_i) \mleft[ \hat S^{(0)}_i \hat \rho (t) \hat S^{(0)}_i - \hat S^{(0)}_i \hat S^{(0)}_i \hat \rho (t)  \mright] \nn \\
&&+ {\rm \Omega}^{'+}_i (T_i) \mleft[ \hat S^{(0)}_i \hat \rho (t) \hat S^{(0)}_i - \hat \rho (t)\hat S^{(0)}_i (\omega') \hat S^{(0)}_i \mright] \nn \\
&&+ {\rm \Omega}^{-}_i (T_i) \mleft[ \hat S^{(0)}_i \hat \rho (t) \hat S^{(0)}_i - \hat \rho (t)\hat S^{(0)}_i \hat S^{(0)}_i  \mright] \nn \\
&&+ {\rm \Omega}^{'-}_i (T_i) \mleft[ \hat S^{(0)}_i \hat \rho (t) \hat S^{(0)}_i - \hat S^{(0)}_i \hat S^{(0)}_i \hat \rho (t) \mright]
%&&+ \Omega (T_i) \lind{\hat S^{(0)}_i} \bigg\}
\label{Lgme1}
\eea
with
\be
{\rm \Omega}_i^{'\pm} (T_i) = \int_0^t d\tau \int_0^\infty d\nu g_i (\nu) \abssq{\alpha_i (\nu)} \mleft[ n (\nu, T_i) + 1 \mright]e^{\pm \imath \nu \tau} \, ,
\ee
\be
{\rm \Omega}_i^{\pm} (T_i) = \int_0^t d\tau \int_0^\infty d\nu g_i (\nu) \abssq{\alpha_i (\nu)} n (\nu, T_i) e^{\pm \imath \nu \tau} \, .
\ee

We also observe that, for the particular case of an Ohmic bath, where
\be
g_i (\nu) \abssq{\alpha_i (\nu)} = \frac{\gamma_i \nu}{2 \pi f_i} \, ,
\ee
with $\gamma_i$ and $f_i$ being, respectively, the damping and the frequency of the considered subsystem, we obtain
\be
\Gamma_i (\omega) = \frac{\gamma_i \omega}{f_i} \,,
\ee
and all the pure dephasing rates give the same result
\be
{\rm \Omega}_i^{'\pm} (T_i) ={\rm \Omega}_i^{\pm} (T_i)={\rm \Omega}(T_i) \,,
\ee
\be
{\rm \Omega}(T_i) = \frac{\gamma_i}{4 f_i} T_i \, .
\ee

In the next section, we apply this generalized dressed master equation to two hybrid quantum systems, comparing the obtained numerical results with previous approaches.

%%%%%%%%%%%%%

\subsubsection{Stability problems}

We observe that the dissipator in \eqref{Lgme1} is not in Lindblad form and, consequently, properties like the positivity of the density matrix and the conservation of the probability cannot be guaranteed. Furthermore, in this framework, some useful theorems~\cite{Kraus2008} on the steady-state behaviour have not been proven yet.

Actually, a careful inspection of \eqref{Lgme1} shows that it can be regarded as approximately Lindblad-like. Specifically, if we consider the interaction picture, each term of \eqref{Lgme1} (except the last) oscillates at frequencies $\pm (\omega - \omega')$. If $(\omega - \omega')$ is significantly larger than the damping rates  $\Gamma_{i}$ of the system, these terms provide negligible contributions when integrating the master equation. Hence, $\abs{\omega - \omega'}$ can be assumed to be of the order of the system linewidths. It is thus reasonable to assume for the thermal populations of the reservoirs $n (\omega, T_i) \simeq n (\omega', T_i)$ and for the dampings $\Gamma_i(\omega)\simeq\Gamma_i(\omega')$. This analysis shows that, within a very good approximation, the dissipator in \eqref{Lgme1} can be regarded to be in Lindblad form.

Although the fast oscillating terms arising in \eqref{Lgme1}, produced from transitions with high frequency differences (not present after the post-trace RWA), should not provide a significant contribution for $\abs{\omega - \omega'} > \Gamma_i$, they can strongly increase the computation time and lead to computational instabilities. In order to overcome these difficulties, we use numerical filtering with a step-like function that sets to zero all the dissipator terms involving frequency differences higher than a certain value $\Lambda$. More specifically, the filtered Liouvillian takes the form
\be
\mathcal{L}_{\rm gme}^{\rm filt} \hat \rho = \mathcal{L}_{\rm gme} \hat \rho \times F(\omega, \omega') \, ,
\ee
where the filter function $F(\omega, \omega')$ can be written in a generalized form as
\be
F (\omega, \omega') = \Theta (\abs{\omega - \omega'}) - \Theta (\abs{\omega - \omega'} -\Lambda) \, ,
\ee
with $\Theta$ the Heaviside step function and $\Lambda$ the bandwidth of the filter. 
%Note that also the additional fast oscillating terms arising when resolving \eqref{liou} are automatically filtered out.
%%%%%%%%%%%%%%%%%%%%%%%%%%%%%%%%%%%%%%%%%%%%%%%%%

\section{Dissipation in the USC regime}

In this section, we apply the generalized master equation presented in the previous section to study the influence of temperature on the dynamics of two open hybrid quantum systems in the USC regime. Specifically, we re-examine the dynamics of the two systems presented in Refs.~\cite{Garziano2015} and \cite{Macri2018}. The first example is a circuit-QED system in the dispersive regime, displaying multiphoton quantum Rabi oscillations. For this setup, we also compare the results obtained with the generalized master equation to those obtained using the dressed approach for anharmonic systems~\cite{Beaudoin2011}.

The second example is an optomechanical system with coexisting harmonic and anharmonic spectra. Specifically, we consider an ultra-high-frequency mechanical oscillator ultrastrongly coupled to a microwave resonator. Very recently, considering zero-temperature reservoirs, it has been shown~\cite{Moore1970} that this system is promising for the observation of the dynamical Casimir effect (DCE), which converts mechanical energy into photon pairs \cite{Nation2012}. Here we analyze the influence of temperature on this fundamental quantum effect. Moreover, in order to understand the impact of the generalized master equation on the dynamics of hybrid quantum systems, we compare the obtained numerical results with those obtained using a previously developed approach for USC optomechanics~\cite{Hu2015}. {Note that all the numerical results are displayed in the lab frame.}

%%%%%%%%%%%%%%%%%%%%%%%%%

\subsection{Circuit QED beyond the RWA}
\label{sec:CircuitQEDBeyondRWA}

In this circuit-QED example, we study a flux qubit coupled to a single-mode resonator~\cite{Garziano2015}. The bare qubit Hamiltonian can be written as
\be
\hat H_q = \omega_q \sz / 2 \, ,
\ee
where the qubit resonance frequency is $\omega_q = \sqrt{\Delta^2 + (2I_p \delta \Phi_x)^2}$, with $\Delta$ the qubit energy gap, $I_p$ the persistent current corresponding to the minima of the qubit potential, and $\delta \Phi_x$ the flux offset. The bare resonator Hamiltonian is
\be
\hat H_c = \omega_c \hat a^\dag \hat a \, ,
\ee
where $\omega_c$ is the frequency of the resonator mode and $\hat a$ ($\hat a^\dag$) is the bosonic annihilation (creation) operator for that mode. The total quantum system is described by the generalized quantum Rabi Hamiltonian
\be
\hat H_S = \hat H_q + \hat H_c + g \hat X \mleft[ \cos (\theta) \sx + \sin (\theta) \sz \mright] \, ,
\label{eq:HGenRabi}
\ee
where the flux dependence is encoded in $\cos (\theta) = \Delta / \omega_q$, $\hat X = \hat a + \hat a^\dag$, and $\sx$, $\sz$ are Pauli matrices.

As shown in Ref.~\cite{Garziano2015}, the lowest energy levels of this system display a well-known avoided level crossing arising for $\omega_q\simeq\omega_c$ (vacuum Rabi splitting). This avoided crossing is due to the coherent coupling of the states $\ket{e, 0}$ and $\ket{g, 1}$, where {\it{g}} ({\it{e}}) indicates the ground (excited) state of the qubit and the second entry in the kets represents the photon number. However, when the RWA breaks down, the counter-rotating terms in \eqref{eq:HGenRabi} must be taken into account, and the total number of excitations in the system is no longer conserved \cite{Niemczyk2010,Sank2016}. As a consequence, the coherent coupling between states with different numbers of excitations, not allowed in the standard Jaynes-Cummings model~\cite{Jaynes1963, Shore1993}, becomes possible through virtual transitions mediated by the counter-rotating terms~\cite{Kockum2017}. This generates several additional avoided level crossings between states with different excitation numbers, e.g., between $\ket{e, 0}$ and $\ket{g, 2}$~\cite{Garziano2015}. 

For our numerical calculations, we consider, as in Ref.~\cite{Garziano2015}, $\omega_c / 2 \pi = \unit[4.0]{GHz}$ and a resonator-qubit coupling strength $g/\omega_c = 0.157$. We focus on the avoided crossing arising at $\omega_q \simeq 2 \omega_c$ between the states $\ket{\psi_\pm} \simeq \frac{1}{\sqrt 2} (\ket{e, 0} \pm \ket{g, 2})$. We set $\omega_q / 2 \pi = 7.97$ GHz (obtained using the qubit parameters $\Delta/h = \unit[2.25]{GHz}$, $2 I_p = \unit[1.97]{nA}$, and $\delta \Phi_x = 3.88 \,\Phi_0$); this is where the splitting reaches its minimum~\cite{Garziano2015}. {The minimum splitting $2 {\rm \Omega}_{\rm eff}$ provides a direct measurement of the effective resonant coupling ${\rm \Omega}_{\rm eff}$ between the states $\ket{e, 0}$ and $\ket{g, 2}$.}

In order to probe this avoided crossing, we consider, as in Ref.~\cite{Garziano2015}, the case where the qubit is directly excited by a Gaussian $\pi$-pulse,
\be
\hat H_p = \mathcal{E} (t) \cos (\omega t) \sx \, ,
\ee
where $\mathcal{E} (t) = {\rm \Omega} \exp [ - (t - t_0)^2 / 2 \tau^2] / (\tau \sqrt{2\pi})$. Here, $\tau$ is the standard deviation and ${\rm \Omega} / \omega_c = (\pi/3) \times 10^{-1}$ the amplitude of the pulse. The center frequency of the pulse corresponds to the middle of the avoided crossing considered here. Specifically, $\omega = (\omega_{3, 0} + \omega_{2, 0}) / 2$, with $\omega_{i, j} = \omega_i - \omega_j$, where we labelled the energy values and the eigenstates of the hybrid system as $\omega_l$ and $\ket{l}$,  with $l = 0, 1, \dots$, such that $\omega_k > \omega_j$ for $k > j$.

The system dynamics is then evaluated using the generalized master equation (gme)
\be
\dot{\hat \rho} = - \imath \comm{\hat H_S + \hat H_p}{\hat \rho} + \mathcal{L}_{\rm gme} \hat \rho \, ,
\ee
where, considering an Ohmic bath, the Liouvillian dissipator can be written as
\bea
{\cal L}_{\rm gme} \hat \rho &=& \sum_{(\omega,\omega') > 0} \frac{1}{2} \Big\{ \frac{\gamma \omega'}{\omega_q} n (\omega', T_\gamma) \mleft[ \hat P^{(-)} (\omega') \hat \rho \hat P^{(+)} (\omega) - \hat P^{(+)} (\omega) \hat P^{(-)} (\omega') \hat \rho \mright] \nn \\
&& + \frac{\gamma \omega}{\omega_q} \mleft[ n (\omega, T_\gamma) + 1 \mright] \mleft[ \hat P^{(+)} (\omega) \hat \rho \hat P^{(-)} (\omega') - \hat P^{(-)} (\omega') \hat P^{(+)} (\omega) \hat \rho \mright] \nn \\
&& + \frac{\gamma \omega}{\omega_q} n (\omega, T_\gamma) \mleft[ \hat P^{(-)} (\omega') \hat \rho \hat P^{(+)} (\omega) - \hat \rho \hat P^{(+)} (\omega) \hat P^{(-)} (\omega') \mright] \nn \\
&& + \frac{\gamma \omega'}{\omega_q} \mleft[ n (\omega', T_\gamma) + 1 \mright] \mleft[ \hat P^{(+)} (\omega) \hat \rho \hat P^{(-)} (\omega') - \hat \rho \hat P^{(-)} (\omega') \hat P^{(+)} (\omega) \mright] \nn \\
&& + \frac{\kappa \omega'}{\omega_c} n (\omega', T_\kappa) \mleft[ \hat A^{(-)}(\omega') \hat \rho \hat A^{(+)} (\omega) - \hat A^{(+)} (\omega) \hat A^{(-)} (\omega') \hat \rho \mright] \nn \\
&& + \frac{\kappa \omega}{\omega_c} \mleft[ n (\omega, T_\kappa) + 1 \mright] \mleft[ \hat A^{(+)} (\omega) \hat \rho \hat A^{(-)} (\omega') - \hat A^{(-)} (\omega') \hat A^{(+)}(\omega) \hat \rho \mright] \nn \\
&& + \frac{\kappa \omega}{\omega_c} n (\omega, T_\kappa) \mleft[ \hat A^{(-)} (\omega') \hat \rho  \hat A^{(+)} (\omega) - \hat \rho \hat A^{(+)} (\omega) \hat A^{(-)} (\omega') \mright] \nn \\
&& + \frac{\kappa \omega'}{\omega_c} \mleft[ n (\omega', T_\kappa) + 1 \mright] \mleft[ \hat A^{(+)} (\omega) \hat \rho \hat A^{(-)} (\omega') - \hat \rho \hat A^{(-)} (\omega') \hat A^{(+)} (\omega) \mright] \Big\} \, .
\eea
Here $\kappa$ and $\gamma$ are the qubit and cavity damping rates, respectively, $\hat A^{(+)}$ and $\hat A^{(-)}$ are the positive- and negative-frequency dressed cavity operators  ($\hat s_i = \hat a$), and $\hat P^{(+)}$ and $\hat P^{(-)}$ are the positive- and negative-frequency dressed qubit operators ($\hat s_i = \sm$). We neglected the very small pure dephasing term in the dissipator [see \eqref{Lgme1}] {and we did not apply any filtering procedure}.

%-----------------------------------------------%
\begin{figure}
\centering
\includegraphics[width = 0.65\linewidth]{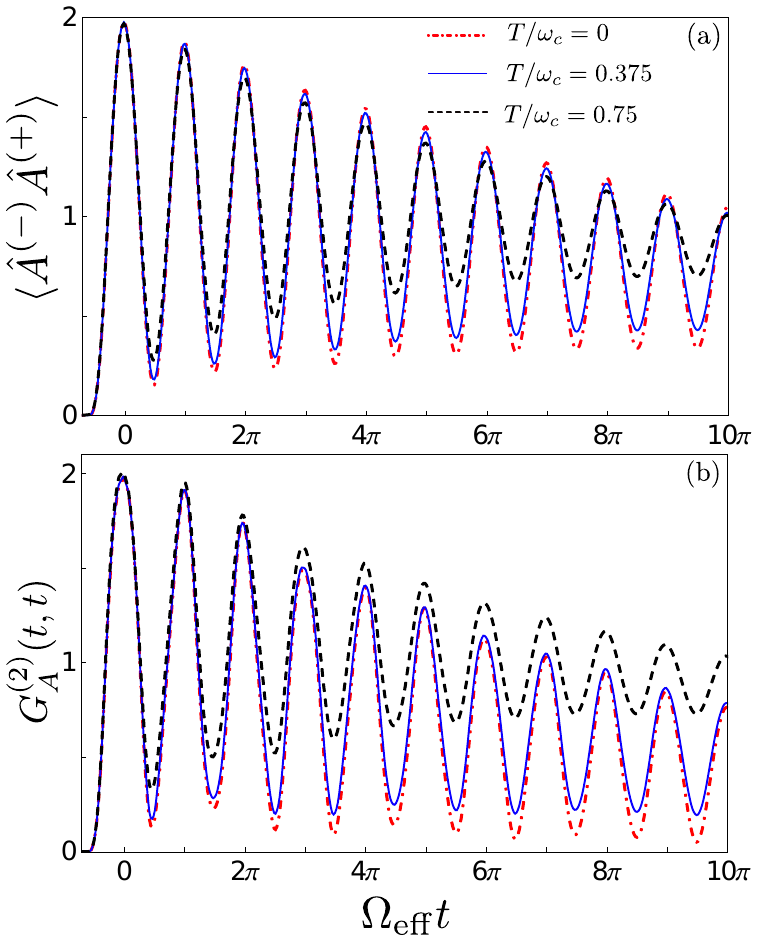}
\caption{{Dynamics of anomalous two-photon vacuum Rabi oscillations.} Results obtained using the generalized-master-equation approach, varying the temperature of both subsystems. (a) Time evolution of the mean cavity photon number $\expec{\hat A^{(-)} \hat A^{(+)}}$ after the arrival of a Gaussian $\pi$-pulse to the qubit. The system starts in the ground state. (b) Two-photon correlation function for the cavity, obtained with the same parameters and conditions. After the arrival of the pulse, independent of the temperature of the reservoirs, the system undergoes vacuum Rabi oscillations showing the reversible exchange of photon pairs between the qubit and the resonator. However, when raising the temperature, due to the increasing decoherence, the oscillations become more damped and the correlation function reaches higher stationary values due to larger incoherent, thermal contributions. Note that the second and the fifth dips are shallower because of some spurious effects generated by other transitions excited by the coherent pulse. All parameters for the simulations are given in the text.  {${\rm \Omega}_{\rm eff}$ on the $x$ axis indicates the effective resonant coupling.}
\label{fig:TempEffectOnMultiPhotonRabi}}
\end{figure}
%-----------------------------------------------%

Figure~\ref{fig:TempEffectOnMultiPhotonRabi} displays the dynamics of the mean cavity photon number $\expec{\hat A^{(-)} \hat A^{(+)}}$ (a) and of the zero-delay two-photon correlation function $G^{(2)}_A (t,t) = \expec{\hat A^{(-)} (t) \hat A^{(-)} (t) \hat A^{(+)} (t) \hat A^{(+)} (t)}$ (b) after the arrival of a Gaussian $\pi$-pulse, evaluated for different temperatures and starting the dynamics with the system in its ground state. We used $T_\gamma / \omega_c = T_\kappa / \omega_c$ and the decoherence rates $\gamma / \omega_c = \kappa / \omega_c = 3.75 \times 10^{-4}$. Note that the output photon flux is proportional to $\expec{\hat A^{(-)} \hat A^{(+)}}$. At $T = 0$ our approach reproduces the two-photon vacuum Rabi oscillations shown in Ref.~\cite{Garziano2015}. Here we study the influence of non-zero temperature on the this anomalous atom-cavity energy exchange. Increasing the temperature, the oscillations become more damped and the energy exchange becomes less effective. This effect is even more pronounced for the two-photon correlation $G^{(2)}_{A}(t,t)$, which displays a stronger thermal sensitivity. These results help to set a limit on the system temperature for the observation of two-photon vacuum Rabi oscillations. 

%-----------------------------------------------%
\begin{figure}
\centering
\includegraphics[width = 0.7\linewidth]{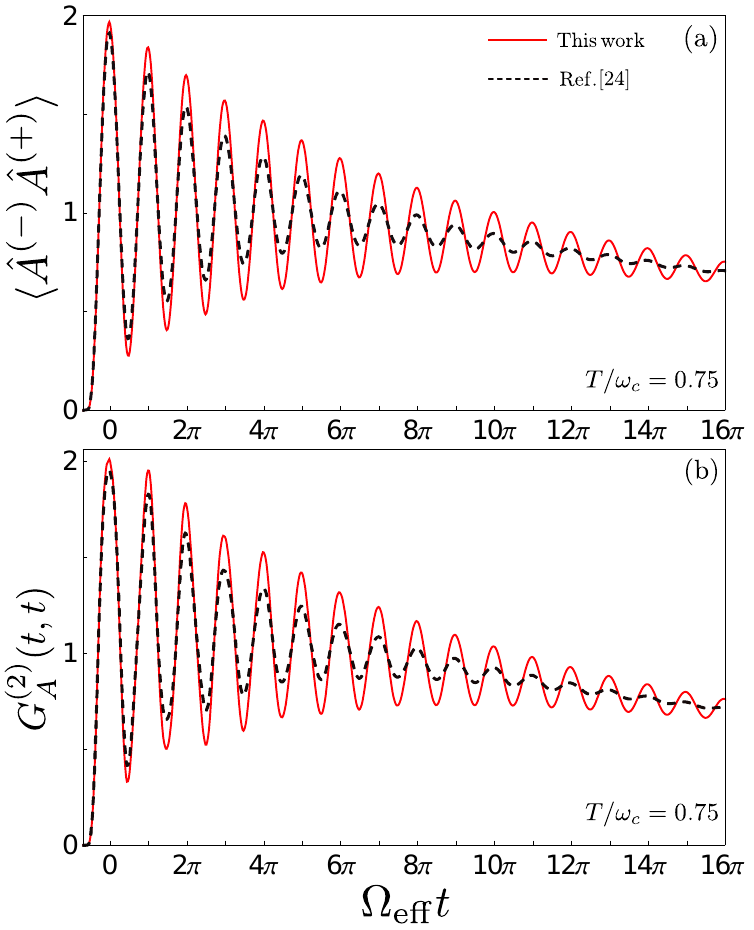}
\caption{Comparison between the results obtained using the generalized-master-equation approach (red solid curves) and the standard dressed master equation (black dashed curves). (a) Time evolution of the mean cavity photon number $\expec{\hat A^{(-)} \hat A^{(+)}}$ at temperature $T / \omega_c = 0.75$ with all other parameters the same as in \figref{fig:TempEffectOnMultiPhotonRabi}. (b) Two-photon correlation functions, obtained with the same parameters and conditions. After the arrival of the pulse, both approaches show the system undergoing two-photon Rabi oscillations and relaxing to thermal equilibrium. However, with the standard dressed master equation, the coherence losses are slightly overestimated (because of the post-trace RWA), so the oscilations are more damped and the stationary value is reached sooner. 
\label{fig:ComparisonBlais}}
\end{figure}
%-----------------------------------------------%

In order to further show the impact of the generalized approach presented in this paper on the dissipative dynamics of cavity-QED systems in the USC regime, we compare the numerical results obtained with the generalized dressed master equation with those obtained using the standard dressed approach of Ref.~\cite{Beaudoin2011}. 

Figure~\ref{fig:ComparisonBlais} shows the mean cavity photon number (a) and the two-photon correlation function (b) evaluated using the generalized dressed master equation (red solid curves) and the standard dressed master equation~\cite{Beaudoin2011} (black dashed curves), calculated with the atom and cavity reservoirs at temperature $T / \omega_c = 0.75$. Both approaches show the system undergoing multiphoton Rabi oscillations and the signals reaching the same stationary values, corresponding to the equilibrium thermal populations. 

We observe that the standard approach overestimates decoherence effects. In the dispersive regime of cavity-QED, pairs of photon-like transitions partially overlap, reducing decoherence effects during the time evolution. This effect is completely neglected in the standard dressed master equation. Further calculations, not shown here, indicate that these discrepancies increase with temperature. These effects lead to an overestimation of the coherence losses of the system, which also can be seen in the behaviour of the two-photon correlation function [\figref{fig:ComparisonBlais}(b)].

It is also important to note that the generalized master equation is able to overcome another limit of validity of the standard dressed approach. As reported in Ref.~\cite{Beaudoin2011}, the standard dressed master equation breaks down in the limit of high excitation numbers, where more transitions might accidentally have the same frequency. The generalized master equation can handle such degenerate transitions well.

%%%%%%%%%%%%%%%%%%%%%%%%%

\subsection{Cavity optomechanics beyond the RWA}
\label{sec:OptomechBeyondRWA}

%%%%%%%%%%%%%

\subsubsection{The full optomechanical Hamiltonian}

In \secref{sec:CircuitQEDBeyondRWA}, we demonstrated that our generalized approach is able to correctly describe systems with quasi-harmonic spectra. In this section, we explore a mixed harmonic-anharmonic behaviour, considering a simple optomechanical system~\cite{Macri2018}, where a single cavity mode of frequency $\omega_c$ is coupled by radiation pressure to a single mechanical mode of a mirror vibrating at frequency $\omega_m$.

Denoting the mechanical bosonic operators $\hat b$, $\hat b^\dag$ and the cavity bosonic operators $\hat a$, $\hat a^\dag$, the system Hamiltonian can be written as~\cite{Law1995}
\be
\hat H_S = \hat H_0 + \hat V_{\rm om} + \hat V_{\rm DCE} \, ,
\label{Hs}
\ee
where
\be
\hat H_0 =  \omega_c \hat a^\dag \hat a + \omega_m \hat b^\dag \hat b
\label{H0}
\ee
is the unperturbed Hamiltonian,
\be
\hat V_{\rm om} = g \hat a^\dag \hat a \mleft( \hat b + \hat b^\dag \mright)
\ee
is the standard optomechanical interaction Hamiltonian, and
\be
\hat V_{\rm DCE} = \frac{g}{2} \mleft( \hat a^2 + \hat a^{\dag 2} \mright) \mleft( \hat b + \hat b^\dag \mright)
\label{VDCE}
\ee
describes the emission of photon pairs induced by the mechanical motion predicted by the DCE~\cite{Johansson2010,Wilson2011,Nation2012}. When treating most optomechanics experiments until now, $\hat V_{\rm DCE}$ has been neglected. This is a very good approximation when the mechanical frequency is much smaller than the cavity frequency (which is the most common experimental situation), because $\hat V_{\rm DCE}$ connects bare states with an energy difference $2 \omega_c \pm \omega_m$ which then is much larger than the coupling strength $g$. With this approximation, the resulting Hamiltonian, $\hat H_0 + \hat V_{\rm om}$, conserves the number of photons and can be analytically diagonalized. However, when considering ultra-high-frequency mechanical oscillators, with resonance frequencies in the GHz spectral range, coupled to a microwave resonator, $\hat V_{\rm DCE}${, which does not conserve the photon number,} cannot be neglected any more~\cite{Macri2018}.

As shown in Ref.~\cite{Macri2018}, such a system displays an energy level spectrum with a ladder of avoided level crossings arising from the coherent coupling induced by $\hat V_{\rm DCE}$ between the states $\ket{n, k_n}$ and $\ket{n+2, (k - q)_{n+2}}$, occurring when the energies of the initial and final states coincide ($2 \omega_c \simeq q \omega_m$). Here the first number in the ket denotes photon number and the second denotes phonon number (with the photon number as a subscript since the photons displace the mechanical Fock state). For example, with $q = 1$, we have the standard resonance condition for the DCE ($2 \omega_c \simeq \omega_m$~\cite{Lambrecht1996}), in which case $\hat V_{\rm DCE}$ gives rise to a resonant coupling between the states $\ket{0, k}$ and $\ket{2, (k - 1)_2}$ with $k \geq 1$, converting a phonon into a photon pair.

When $\hat V_{\rm DCE}$ is taken into account, the system Hamiltonian does not conserve the number of photons (the phonon number is not conserved even in the standard optomechanical Hamiltonian). For example, the ground state of $\hat H_S$ contains photons, i.e., $\brakket{E_0}{\hat a^\dag a}{E_0} \neq 0$. Therefore, in analogy to USC cavity QED, a careful treatment of dissipation and input-output theory is required. If the standard photon and phonon operators were used to describe the interaction with the outside world, unphysical effects would arise.

%%%%%%%%%%%%%

\subsubsection{The impact of temperature on the dynamical Casimir effect}

It has been shown~\cite{Macri2018} that this system can be used to demonstrate the conversion of mechanical energy into photon pairs (DCE). The calculations in Ref.~\cite{Macri2018} were performed using a dressed master equation without the post-trace RWA, developed only for the case of zero-temperature reservoirs. Here we instead apply the generalized master equation presented in \secref{sec:GeneralizedME}, in order to study the influence of temperature on the energy conversion from phonons to photons.

For our numerical calculation we consider a normalized optomechanical coupling $g / \omega_m = 0.1$, a mechanical damping rate $\gamma / \omega_m = 0.05$, and a cavity damping rate $\kappa = \gamma / 2$. We focus on the avoided level crossing between the states $\ket{0, 2}$ and $\ket{2, 0_2}$ at $\omega_m \simeq \omega_c$. We consider the resonant condition, corresponding to the minimum level splitting: $\omega_c / \omega_m = 1.016$. 

As in Ref.~\cite{Macri2018}, we consider a continuous coherent drive of the mechanical oscillator,
\be
\hat H_d = {\rm \Omega} \mleft( \hat b e^{-\imath \omega_m t} + \hat b^\dag e^{\imath \omega_m t} \mright) \, ,
\ee
with frequency resonant with the oscillating mirror and amplitude ${\rm \Omega} = \gamma / 2$. The dynamics giving rise to the DCE is then described by the filtered generalized master equation ($\Lambda=10\gamma$)
\be
\dot{\hat \rho} = - i \comm{\hat H_S + \hat H_d}{\hat \rho} + \mathcal{L}^{\rm filt}_{\rm gme} \hat \rho \, ,
\ee
where the Liouvillian superoperator can be written as
\bea
\mathcal{L}^{\rm filter}_{\rm gme} \hat \rho &=& \sum_{(\omega,\omega') > 0} \frac{1}{2} \Big\{ \gamma n (\omega', T_\gamma) \mleft[ \hat B^{(-)} (\omega') \hat \rho \hat B^{(+)} (\omega) - \hat B^{(+)} (\omega) \hat B^{(-)} (\omega') \hat \rho \mright] \nn \\
&& + \gamma \mleft[ n (\omega, T_\gamma) + 1 \mright] \mleft[ \hat B^{(+)} (\omega) \hat \rho \hat B^{(-)} (\omega') - \hat B^{(-)} (\omega') \hat B^{(+)}(\omega) \hat \rho \mright] \nn \\
&& + \gamma n(\omega, T_\gamma) \mleft[ \hat B^{(-)} (\omega') \hat \rho \hat B^{(+)} (\omega) - \hat \rho \hat B^{(+)} (\omega) \hat B^{(-)} (\omega') \mright] \nn \\
&& + \gamma \mleft[ n (\omega', T_\gamma) + 1 \mright] \mleft[ \hat B^{(+)} (\omega) \hat \rho \hat B^{(-)} (\omega') - \hat \rho \hat B^{(-)} (\omega') \hat B^{(+)} (\omega) \mright] \nn \\
&& + \kappa n (\omega', T_\kappa) \mleft[ \hat A^{(-)} (\omega') \hat \rho \hat A^{(+)} (\omega) - \hat A^{(+)} (\omega) \hat A^{(-)} (\omega') \hat \rho \mright] \nn \\
&& + \kappa \mleft[ n (\omega, T_\kappa) + 1 \mright] \mleft[ \hat A^{(+)} (\omega) \hat \rho \hat A^{(-)} (\omega') - \hat A^{(-)} (\omega') \hat A^{(+)} (\omega) \hat \rho \mright] \nn \\
&& + \kappa n(\omega, T_\kappa) \mleft[ \hat A^{(-)}(\omega') \hat \rho \hat A^{(+)} (\omega) - \hat \rho \hat A^{(+)} (\omega) \hat A^{(-)} (\omega') \mright] \nn \\
&& + \kappa \mleft[ n (\omega', T_\kappa) + 1 \mright] \mleft[ \hat A^{(+)} (\omega) \hat \rho \hat A^{(-)} (\omega') - \hat \rho \hat A^{(-)} (\omega') \hat A^{(+)} (\omega) \mright] \Big\}\times F (\omega, \omega') \, ,
\eea
where $\hat A^{(+)}$ and $\hat A^{(-)}$ are the positive- and negative-frequency dressed cavity operators ($\hat s_i = \hat a$), and $\hat B^{(+)}$ and $\hat B^{(-)}$ are the positive- and negative-frequency dressed mechanical operators ($\hat s_i = \hat b$).

%-----------------------------------------------%
\begin{figure}
\centering
\includegraphics[width = \linewidth]{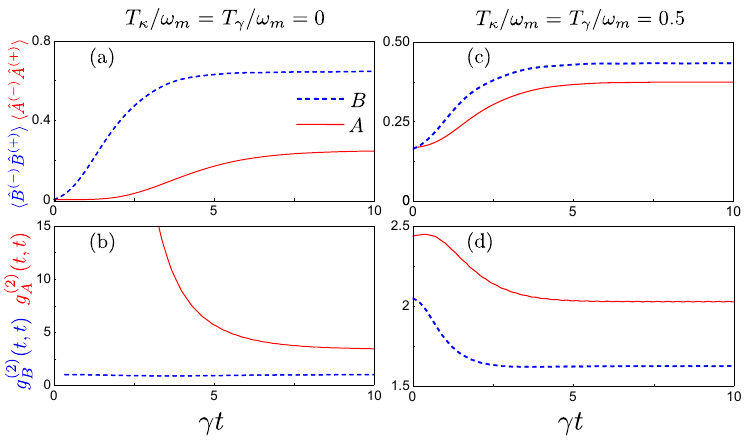}
\caption{Results for the DCE at different temperatures, obtained using the generalized-master-equation approach. (a, b) System dynamics for $\omega_c \simeq \omega_m$, under coherent mechanical pumping, in perfect cooling conditions $T_\gamma = T_{\kappa} = 0$, starting the dynamics from the ground state. (c, d) The same, but with $T_\gamma / \omega_m = T_\kappa /\omega_m = 0.5$ and the initial state being the thermal state with $T / \omega_m = 0.5$. The blue dashed curves show the mean phonon number $\expec{\hat B^{(-)} \hat B^{(+)}}$ in (a, c) and the phonon-phonon correlation function $g_B^{(2)} (t, t)$ in (b, d). The red solid curves describe the mean cavity photon number $\expec{\hat A^{(-)} \hat A^{(+)}}$ in (a, c) and the zero-delay normalized photon-photon correlation function $g^{(2)}_A (t, t)$ in (b, d). All parameters for the simulations are given in the text.
\label{fig:DCETemps}}
\end{figure}
%-----------------------------------------------%

In \figref{fig:DCETemps}, we show the photonic and phononic populations, $\expec{\hat A^{(-)} \hat A^{(+)}}$ and $\expec{\hat B^{(-)} \hat B^{(+)}}$, and the relative two-photon and two-phonon correlation functions,
\bea
g^{(2)}_A (t, t) &=& \frac{\expec{\hat A^{(-)} (t) \hat A^{(-)}(t) \hat A^{(+)}(t) \hat A^{(+)}(t)}}{\expec{\hat A^{(-)} (t) \hat A^{(+)} (t)}^2} \, , \\ 
g^{(2)}_B (t, t) &=& \frac{\expec{\hat B^{(-)} (t) \hat B^{(-)} (t) \hat B^{(+)} (t) \hat B^{(+)} (t)}}{\expec{\hat B^{(-)} (t) \hat B^{(+)} (t)}^2} \, .
\eea
Figures~\ref{fig:DCETemps}(a, b) display the results of calculations done with zero-temperature reservoirs for both subsystems and starting the dynamics from the ground state. Figures~\ref{fig:DCETemps}(c, d) display the results of calculations for reservoirs with $T_\gamma / \omega_m = T_\kappa/\omega_m = 0.5$ and with the system initially in  thermal equilibrium with those reservoirs.

At $T = 0$, with the system starting in its ground state, the photonic and phononic populations start from zero and, due to the coherent pumping, reach non-zero stationary values. The photonic correlation function $g_A^{(2)} (t, t)$ is initially much higher than two, suggesting photon-pair emission. As time goes on, $g_A^{(2)} (t, t)$ decreases significantly due to losses which affect the photon-photon correlations, and also due to the increase of the mean photon number (note that $g_A^{(2)} (t, t)$, owing to the squared denominator, is an intensity-dependent quantity). The mechanical correlation function $g_B^{(2)} (t, t)$, on the contrary, has an almost constant value ($g_B^{(2)} (t, t) \approx 1$), showing that the mechanical system is mainly in the coherent state produced by the pumping. 

For reservoirs with non-zero temperature, the phonon and photon populations, starting from their thermal-equilibrium values, equilibrate to lower steady-state values. This reduction of both populations originates from the increase of the decay rate of the coherent contributions with increasing temperature. We also note that the difference between the two steady-state values is reduced at higher temperatures, due to the thermal contributions. At $T \neq 0$, a fraction of the observed photons, as expected, does not come from the mechanical-to-optical energy conversion, but, trivially, from the photonic thermal reservoir.
  
This picture is confirmed by comparing the dynamics of the higher-order correlation functions [Figs.~\ref{fig:DCETemps}(c, d)]. Specifically, at higher temperature, we observe a strong decrease of $g_A^{(2)} (t, t)$, showing that a reduced fraction of photons is emitted in pairs. However, the photon-photon correlation functions remains, even in the steady state, {\it higher} than the thermal value $g_A^{(2)} (t, t) = 2$. The phonon-phonon correlation starts from a value $\simeq 2$ corresponding to the initial incoherent thermal state and, as time goes on, decays to a stationary value higher than one due to the incoherent thermal excitations provided by the interaction with the thermal reservoirs. 

Furthermore, in \figref{fig:DCETemps}(d), the photon-photon and the phonon-phonon correlation functions do not start from the same initial value. This effect is due to the $\hat V_{\rm DCE}$ term which, owing to its non-bilinear form, modifies the thermodynamic equilibrium of the initial state of the system. The $\hat V_{\rm DCE}$ contribution leads to a separation of the correlation-function values with size proportional to the temperature. This separation thus vanishes trivially for $T = 0$, when the $\hat V_{\rm DCE}$ term becomes negligible. 

The results obtained clearly show that the generalized dressed master equation provided here is able to describe dissipation in hybrid quantum systems with coexisting coherent phases (provided, e.g., by means of a continuous drive) and incoherent phases (provided, e.g., by thermal reservoirs or thermal-like pumping). The behaviour of the one- and two-photon correlation functions show that signatures of the DCE can be observed even in the presence of a non-negligible amount of thermal noise. It thus demonstrates that this effect can be observed in a real experimental set-up, where perfect cooling conditions cannot be reached. Although the number of Casimir photon pairs produced depends on the thermal noise injected into the system, our results here show that the DCE remains detectable even at relatively high temperatures.

%%%%%%%%%%%%%

\subsubsection{Comparison to other approaches}

%-----------------------------------------------%
\begin{figure}
\centering
\includegraphics[width = 0.7\linewidth]{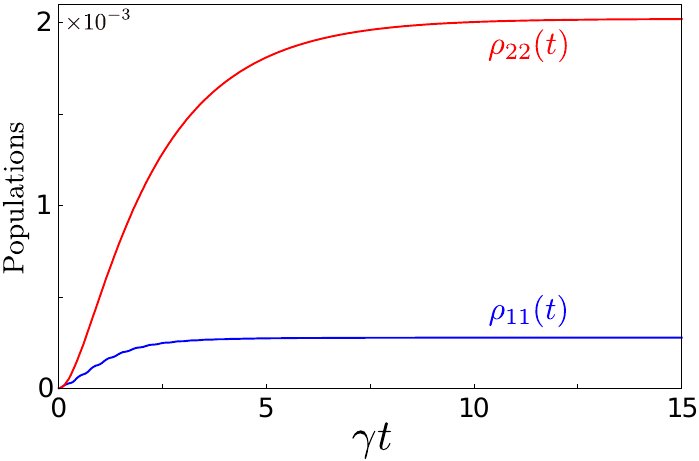}
\caption{State populations obtained using the master-equation approach of Ref.~\cite{Hu2015}. The red (blue) solid curve shows the time evolution of the population of the one-photon state $\ket{1, 0}$ (the one-phonon state $\ket{0, 1}$) labelled $\rho_{22} (t)$ [$\rho_{11} (t)$] under perfect cooling conditions $T_\gamma = T_\kappa = 0$ and without any pumping. The initial state is the ground state $\ket{0, 0}$. All other parameters are the same as in \figref{fig:DCETemps}. In these conditions, without any external driving or thermal excitations, the system is expected to remain in the ground state. However, the plot clearly shows a non-zero population in both the one-photon and one-phonon states. This indicates that this approach is {\it not} able to correctly describe optomechanical systems when the $\hat V_{\rm DCE}$ contribution no longer can be neglected.
\label{fig:TestHu}}
\end{figure}
%-----------------------------------------------%

As already mentioned in the introduction, and demonstrated in Ref.~\cite{Beaudoin2011}, the use of a master equation with a dissipator not taking into account the interaction between the subsystems can lead to unphysical results. Hu {\em et al.}~derived~\cite{Hu2015} a dressed master equation specifically developed to describe dissipation in optomechanical systems characterized by the standard optomechanical Hamiltonian, $\hat H_S = \hat H_0 + \hat V_{\rm om}$, in the USC regime. Here we show that this master equation {\it fails} when considering the complete optomechanical Hamiltonian $\hat H_S = \hat H_0 + \hat V_{\rm om} + \hat V_{\rm DCE}$.

In \figref{fig:TestHu}, we display results obtained describing the dynamics of our optomechanical system in perfect cooling conditions, without any pumping, with the master equation provided in Ref.~\cite{Hu2015}, including the $\hat V_{\rm DCE}$ term as a perturbation in the dynamics. In these conditions, evaluating the dynamics with the system initially in the ground state (an eigenstate of the system), zero population is expected in the states with one photon, $\ket{1,0}$, and one phonon, $\ket{0,1}$. However, \figref{fig:TestHu} clearly shows non-zero populations. This anomalous effect occurs because, due to the additional $\hat V_{\rm DCE}$ term, the number of photons is no longer conserved and consequently the eigenstates of the Hamiltonian changes. In this case, the master equation provided in Ref.~\cite{Hu2015} does not describe interactions between subsystems and reservoirs in terms of the correct eigenstates, which leads to an unphysical evolution of the initial ground state. This result shows once more the importance of expressing the system operators in the basis of the system eigenstates when describing interactions with reservoirs to derive a correct master equation.

%%%%%%%%%%%%%%%%%%%%%%%%%%%%%%%%%%%%%%%%%%%%%%%%%

\section{Conclusions}
\label{sec:Conclusions}

We have presented a generalized dressed master equation, valid for arbitrary open hybrid quantum system interacting with thermal reservoirs and for arbitrary strength of the coupling between the components of the hybrid system. Our approach was derived within the Born-Markov approximation, including the pure dephasing terms and without performing the usual post-trace RWA. Therefore, our approach is able to handle dynamics in systems with both harmonic, quasi-harmonic, and anharmonic transitions. Moreover, this approach is not limited to systems displaying parity symmetry. Unfortunately the dissipator obtained includes rapidly oscillating terms that can cause numerical instabilities. In order to fix this problem, we introduced a filtering procedure which eliminates the fast-oscillating terms which do not contribute to the coarse-grained dynamics. This filtering has the added benefit of reducing computation times.

We applied our generalized approach to study the influence of temperature on multiphoton vacuum Rabi oscillation in a circuit-QED system in the dispersive regime. We compared our results with those obtained using the dressed master equation of Ref.~\cite{Beaudoin2011}. We found that both approaches describe multiphoton Rabi oscillations and reach the same stationary state (the thermal equilibrium). However, the standard master equation overestimates decoherence effects since it does not take into account the partial overlap of photon-like transitions, which reduces the decoherence during the time evolution.

We also studied the influence of temperature on the conversion of mechanical energy into photon pairs (DCE) in an optomechanical system, recently described in Ref.~\cite{Macri2018} for zero-temperature reservoirs. In this case, we showed that the DCE can be observed also in the presence of a significant amount of thermal noise.

Finally, we demonstrated that the master-equation approach provided in Ref.~\cite{Hu2015} for optomechanical systems with ultrastrong coupling fails when considering the full optomechanical Hamiltonian including the $\hat V_{\rm DCE}$ term. Specifically, under these conditions, the master equation provided in Ref.~\cite{Hu2015} does not describe interactions between the components and reservoirs correctly in terms of transitions between eigenstates of the hybrid system. Because of this shortcoming, that approach leads to an unphysical evolution of the initial ground state to excited states even at zero temperature and without any external pumping. This example clearly shows that the general master-equation approach provided here is necessary to describe dissipation of general open hybrid quantum systems interacting with thermal reservoirs. 

\section{Acknowledgements}

F.N. is supported in part by the: 
MURI Center for Dynamic Magneto-Optics via the 
Air Force Office of Scientific Research (AFOSR) (FA9550-14-1-0040), 
Army Research Office (ARO) (\textcolor{red}{Grant No. W911NF-18-1-0358}), 
Asian Office of Aerospace Research and Development (AOARD) (Grant No. FA2386-18-1-4045), 
Japan Science and Technology Agency (JST) (\textcolor{red}{Q-LEAP program, the ImPACT program and CREST Grant No. JP-MJCR1676), Japan Society for the Promotion of Science (JSPS) (JSPS-RFBR Grant No. 17-52-50023 and JSPS-FWO Grant No. VS.059.18N), RIKEN-AIST Challenge Research Fund, and the John Templeton Foundation.} 

\appendix

\section{Derivation of the dressed master equation}
\label{app:DerivationDressedME}

Starting from the system-bath Hamiltonian in \eqref{HSB1}, and following the standard procedure~\cite{Breuer2002}, i.e., performing the second-order Born approximation, the Markov approximation, and considering reservoirs with a continuum of frequencies, we obtain
\bea
\dot{\hat{\tilde \rho}} (t) &=& \int_0^t dt' \mleft[ \hat{\tilde S}_i(t') \hat{\tilde \rho} (t') \hat{\tilde S}_i (t) - \hat{\tilde S}_i (t) \hat{\tilde S}_i (t') \hat{\tilde \rho} (t') \mright] \expec{\hat{\tilde B}_i^\dag (t) \hat{\tilde B}_i (t')} \nn \\
&&+ \int_0^t dt' \mleft[ \hat{\tilde S}_i (t) \hat{\tilde \rho} (t) \hat{\tilde S}_i (t') - \hat{\tilde \rho} (t') \hat{\tilde S}_i (t') \hat{\tilde S}_i (t) \mright] \expec{\hat{\tilde B}_i^\dag (t') \hat{\tilde B}_i (t)} \nn \\ 
&&+ \int_0^t dt' \mleft[ \hat{\tilde S}_i(t') \hat{\tilde \rho} (t) \hat{\tilde S}_i(t) - \hat{\tilde S}_i(t) \hat{\tilde S}_i(t') \hat{\tilde \rho} (t') \mright] \expec{\hat{\tilde B}_i(t) \hat{\tilde B}^\dag_i (t')} \nn \\ 
&&+\int_0^t dt' \mleft[ \hat{\tilde S}_i(t) \hat{\tilde \rho} (t') \hat{\tilde S}_i(t') - \hat{\tilde \rho} (t') \hat{\tilde S}_i(t') \hat{\tilde S}_i(t) \mright] \expec{\hat{\tilde B}_i (t') \hat{\tilde B}^\dag_i (t)} \, ,
\label{rhop}
\eea
where $\expec{\hat{\tilde B}_i^\dag (t) \hat{\tilde B}_i (t')}$ and $\expec{\hat{\tilde B}_i^\dag (t') \hat{\tilde B}_i (t)}$ are the reservoir correlation functions
\bea
\expec{\hat{\tilde B}_i^\dag (t) \hat{\tilde B}_i (t')} &=& \int_0^\infty d\nu g_i (\nu) \abssq{\alpha_i (\nu)} n (\nu, T_i) e^{\imath \nu (t - t')} \, ,
\label{b1} \\
\expec{\hat{\tilde B}_i (t) \hat{\tilde B}^\dag_i (t')} &=& \int_0^\infty d\nu g_i (\nu) \abssq{\alpha_i (\nu)} \mleft[ n (\nu, T_i) + 1 \mright] e^{-\imath \nu (t - t')} \, ,
\label{b2}
\eea
with $g (\nu)$ being the reservoir density of states and $\alpha (\nu)$ the system-reservoir coupling strength. Substituting Eqs.~(\ref{b1}) and (\ref{b2}) into \eqref{rhop}, and performing the change of variable $\tau = t - t'$, we obtain
\be
\dot{\hat{\tilde \rho}} (t) = \sum_i \sum_{ \omega, \omega'} \mleft[ \hat A^i_{\omega, \omega'} (t) + \hat B^i_{\omega, \omega'} (t) + \hat C^i_{\omega, \omega'} (t) + \hat D^i_{\omega, \omega'} (t) \mright] \, ,
\label{rhop1}
\ee
where
\bea
\hat A^i_{\omega, \omega'} (t) &=& \int_0^t d\tau e^{-\imath (\omega + \omega') t} e^{\imath \omega' \tau} \mleft[ \hat S_i (\omega') \hat{\tilde \rho} (t) \hat S_i (\omega) - \hat S_i (\omega) \hat S_i (\omega') \hat{\tilde \rho} (t) \mright] \nn \\
&&\times \int_0^\infty d\nu g_i (\nu) \abssq{\alpha_i (\nu)} n (\nu, T_i) e^{\imath \nu \tau} \, ,\nn \\
\hat B^i_{\omega, \omega'} (t) &=& \int_0^t d\tau e^{-\imath (\omega + \omega') t} e^{\imath \omega \tau} \mleft[ \hat S_i (\omega') \hat{\tilde \rho} (t) \hat S_i (\omega) - \hat{\tilde \rho} (t) \hat S_i (\omega) \hat S_i (\omega') \mright] \nn \\
&&\times \int_0^\infty d\nu g_i (\nu) \abssq{\alpha_i (\nu)} n (\nu, T_i) e^{-\imath \nu \tau} \, , \\
\hat C^i_{\omega, \omega'} (t) &=& \int_0^t d\tau e^{-\imath (\omega + \omega') t} e^{\imath \omega \tau} \mleft[ \hat S_i (\omega) \hat{\tilde \rho} (t) \hat S_i (\omega') - \hat S_i (\omega') \hat S_i (\omega) \hat{\tilde \rho} (t) \mright] \nn \\
&&\times \int_0^\infty d\nu g_i (\nu) \abssq{\alpha_i (\nu)} \mleft[ n (\nu, T_i) + 1 \mright] e^{-\imath \nu \tau} \, , \nn \\
\hat D^i_{\omega, \omega'} (t) &=& \int_0^t d\tau e^{-\imath (\omega + \omega') t} e^{\imath \omega' \tau} \mleft[ \hat S_i (\omega) \hat{\tilde \rho} (t) \hat S_i (\omega') - \hat{\tilde \rho} (t) \hat S_i (\omega') \hat S_i (\omega) \mright] \nn \\
&&\times \int_0^\infty d\nu g_i (\nu) \abssq{\alpha_i (\nu)} \mleft[ n (\nu, T_i) + 1 \mright] e^{\imath \nu \tau} \nn \, ,
\eea
Assuming that the integrands decay on a much shorter time scale than that of the reservoir correlation functions, we can extend the $\tau$ integration to infinity. Evaluating both the integrals without performing any approximation except for the Born-Markov approximation, the master equation in the Schr\"odinger picture can be written
\be
\dot{\hat \rho} = - i \comm{\hat H_S}{\hat \rho} + \mathcal{L}_{\rm gme} \hat \rho \, ,
\ee
with the Lindbladian superoperator that in the most general form can be written as,
\bea
\mathcal{L}_{\rm gme} \hat \rho &=& \frac{1}{2} \sum_i \sum_{\omega, \omega'} \bigg\{ \Gamma_i (-\omega') n (-\omega', T_i) \mleft[ \hat S_i (\omega') \hat{\tilde \rho} (t) \hat S_i (\omega) - \hat S_i (\omega) \hat S_i (\omega') \hat \rho (t)\mright] \nn \\
&&+ \Gamma_i (\omega) n (\omega, T_i) \mleft[ \hat S_i (\omega') \hat \rho (t) \hat S_i (\omega) - \hat \rho (t) \hat S_i (\omega) \hat S_i (\omega')  \mright] \nn \\
&&+ \Gamma_i (\omega) [n (\omega, T_i) + 1] \mleft[ \hat S_i (\omega) \hat \rho (t) \hat S_i (\omega') - \hat S_i (\omega') \hat S_i (\omega) \hat \rho (t) \mright] \nn \\
&&+ \Gamma_i (-\omega') [n (-\omega', T_i) + 1] \mleft[ \hat S_i (\omega) \hat \rho (t) \hat S_i (\omega') - \hat \rho (t) \hat S_i (\omega') \hat S_i (\omega) \mright] \bigg\}\,.
\label{Lgme}
\eea
Both $\Gamma_i(\omega)$ and $n(\omega,T_i)$ are non zero only for $\omega>0$ thus, using the definitions in Eq.~(\ref{eq:SPlus}), \eqref{Lgme} can be written as
\bea
{\cal L}_{\rm gme} \hat \rho &=& \frac{1}{2} \sum_i \sum_{(\omega, \omega') > 0} \bigg\{ \Gamma_i (\omega') n (\omega', T_i) \mleft[ \hat S^{(-)}_i (\omega') \hat \rho (t) \hat S^{(+)}_i (\omega) - \hat S^{(+)}_i (\omega) \hat S^{(-)}_i (\omega') \hat \rho (t) \mright] \nn \\
&&+ \Gamma_i (\omega) n (\omega, T_i) \mleft[ \hat S^{(-)}_i (\omega') \hat \rho (t) \hat S^{(+)}_i (\omega) - \hat \rho (t) \hat S^{(+)}_i (\omega) \hat S^{(-)}_i (\omega') \mright] \nn \\
&&+ \Gamma_i (\omega) [n (\omega, T_i) + 1] \mleft[ \hat S^{(+)}_i (\omega) \hat \rho (t) \hat S^{(-)}_i (\omega') - \hat S^{(-)}_i (\omega') \hat S^{(+)}_i (\omega) \hat \rho (t) \mright] \nn \\
&&+ \Gamma_i (\omega') [n (\omega', T_i) + 1] \mleft[ \hat S^{(+)}_i (\omega) \hat \rho (t) \hat S^{(-)}_i (\omega') - \hat \rho (t) \hat S^{(-)}_i (\omega') \hat S^{(+)}_i (\omega) \mright] \nn \\
&&+ \Gamma_i (\omega') n (\omega', T_i) \mleft[ \hat S^{(-)}_i (\omega') \hat \rho (t) \hat S^{(-)}_i (\omega) - \hat S^{(-)}_i (\omega) \hat S^{(-)}_i (\omega') \hat \rho (t) \mright] \nn \\
&&+ \Gamma_i (\omega') [n (\omega', T_i) + 1] \mleft[ \hat S^{(-)}_i (\omega) \hat \rho (t) \hat S^{(-)}_i (\omega') - \hat  \rho (t) \hat S^{(-)}_i (\omega') \hat S^{(-)}_i (\omega) \mright] \nn \\
&&+ \Gamma_i (\omega) n (\omega, T_i) \mleft[ \hat S^{(+)}_i (\omega') \hat \rho (t) \hat S^{(+)}_i (\omega) - \hat \rho (t)\hat S^{(+)}_i (\omega) \hat S^{(+)}_i (\omega')  \mright] \nn \\
&&+ \Gamma_i (\omega) [n (\omega, T_i) + 1] \mleft[ \hat S^{(+)}_i (\omega) \hat \rho (t) \hat S^{(+)}_i (\omega') - \hat S^{(+)}_i (\omega') \hat S^{(+)}_i (\omega) \hat \rho (t) \mright] \nn \\
&&+ \Gamma_i (\omega') n (\omega', T_i) \mleft[ \hat S^{(-)}_i (\omega') \hat \rho (t) \hat S^{(0)}_i - \hat S^{(0)}_i \hat S^{(-)}_i (\omega') \hat \rho (t)  \mright]\nn \\
&&+ \Gamma_i (\omega') [n (\omega', T_i) + 1] \mleft[ \hat S^{(0)}_i \hat \rho (t) \hat S^{(-)}_i (\omega') - \hat \rho (t)\hat S^{(-)}_i (\omega') \hat S^{(0)}_i \mright] \nn \\
&&+ \Gamma_i (\omega) n (\omega, T_i) \mleft[ \hat S^{(0)}_i \hat \rho (t) \hat S^{(+)}_i (\omega) - \hat \rho (t)\hat S^{(+)}_i (\omega) \hat S^{(0)}_i  \mright] \nn \\
&&+ \Gamma_i (\omega) [n (\omega, T_i) + 1] \mleft[ \hat S^{(+)}_i (\omega) \hat \rho (t) \hat S^{(0)}_i - \hat S^{(0)}_i \hat S^{(+)}_i (\omega) \hat \rho (t) \mright] \nn \\
&&+ {\rm \Omega}^{+}_i (T_i) \mleft[ \hat S^{(0)}_i \hat \rho (t) \hat S^{(0)}_i - \hat S^{(0)}_i \hat S^{(0)}_i \hat \rho (t)  \mright] \nn \\
&&+ {\rm \Omega}^{'+}_i (T_i) \mleft[ \hat S^{(0)}_i \hat \rho (t) \hat S^{(0)}_i - \hat \rho (t)\hat S^{(0)}_i (\omega') \hat S^{(0)}_i \mright] \nn \\
&&+{\rm \Omega}^{-}_i (T_i) \mleft[ \hat S^{(0)}_i \hat \rho (t) \hat S^{(0)}_i - \hat \rho (t)\hat S^{(0)}_i \hat S^{(0)}_i  \mright] \nn \\
&&+ {\rm \Omega}^{'-}_i (T_i) \mleft[ \hat S^{(0)}_i \hat \rho (t) \hat S^{(0)}_i - \hat S^{(0)}_i \hat S^{(0)}_i \hat \rho (t) \mright]
%&&+ \sum_i \Omega (T_i) \lind{\hat S^{(0)}_i} \bigg\}
\label{lgmerewritten}
\eea
with thermal populations
\be
n (\omega, T_i) = \mleft[ \exp{\{\omega / T_i\}} - 1 \mright]^{-1} \, ,
\label{nterm}
\ee
damping rates
\be
\Gamma_i (\omega) = 2 \pi g_i (\omega) \abssq{\alpha_i (\omega)} \, .
\ee
and pure dephasing damping rates
\be
{\rm \Omega}_i^{'\pm} (T_i) = \int_0^t d\tau \int_0^\infty d\nu g_i (\nu) \abssq{\alpha_i (\nu)} \mleft[ n (\nu, T_i) + 1 \mright]e^{\pm \imath \nu \tau} \, ,
\ee
\be
{\rm \Omega}_i^{\pm} (T_i) = \int_0^t d\tau \int_0^\infty d\nu g_i (\nu) \abssq{\alpha_i (\nu)} n (\nu, T_i) e^{\pm \imath \nu \tau} \, .
\ee

%\be
%\Omega (T_i) = 2 \int_0^t d\tau \int_0^\infty d\nu g_i (\nu) \abssq{\alpha_i (\nu)} \mleft[ 2 n (\nu, T_i) + 1 \mright]e^{\pm i \nu \tau} \, .
%\ee
Specifically, the terms in the first four lines of \eqref{lgmerewritten} oscillate at frequencies $\pm (\omega - \omega')$. If $(\omega - \omega')$ is significantly larger than the damping rates $\Gamma_i$ of the system, these terms provide negligible contributions when integrating the master equation. In the generalized approach, these terms are then eliminated by the numerical filtering. The terms in the next four lines of \eqref{lgmerewritten} oscillate at $\pm(\omega'+\omega)$. These terms are clearly rapidly oscillating and thus provide negligible contributions. The terms in the following four lines, oscillating at $+ \omega$, $-\omega'$, are fast-oscillating when considering systems displaying well-separated energy levels with $\omega \gg \Gamma_i$  and, in these cases, can be neglected. Finally, the term in the last four lines arise from degenerate transitions and describe pure dephasing. The contribution of these terms becomes negligible at very low temperatures in the particular case of Ohmic baths. 
Furthermore, it is important to note that, applying the post-trace RWA without considering any parity symmetry of the system, \eqref{Lgme} can be rewritten in a form equal to the standard dressed master equation as in Ref.~\cite{Beaudoin2011}, with a few additional terms provided by the zero-frequency operators $\hat S^{(0)}_i\neq 0$.

\bibliography{refME}
\end{document}